# Interband transitions in semi-metals, semiconductors, and topological insulators: A new driving force for plasmonics and nanophotonics

Johann Toudert,[1,2,*] Rosalía Serna[1]

[1]*Laser Processing Group (LPG), Instituto de Óptica, CSIC, Madrid, Spain*
[2]*Currently with ICFO – Institut de Ciencies Fotoniques, Barcelona Institute of Technology, 08860 Castelldefels (Barcelona), Spain*
*\*johann.toudert@gmail.com*

Plasmonic and Mie resonances in subwavelength nanostructures provide an efficient way to manipulate light below the diffraction limit that has fostered the growth of plasmonics and nanophotonics. Plasmonic resonances have been mainly related with the excitation of free charge carriers, initially in metals, and Mie resonances have been identified in Si nanostructures. Remarkably, although much less studied, semi-metals, semiconductors and topological insulators of the p-block enable plasmonic resonances without free charge carriers *and* Mie resonances with enhanced properties compared with Si. In this review, we explain how interband transitions in these materials show a major role in this duality. We evaluate the plasmonic and Mie performance of nanostructures made of relevant p-block elements and compounds, especially Bi, and discuss their promising potential for applications ranging from switchable plasmonics and nanophotonics to energy conversion, especially photocatalysis.

## 1. Introduction

Optically resonant subwavelength nanostructures have attracted the attention of a broad community of scientists due their excellent capability for the manipulation of light below the diffraction limit, including its harvesting with a high extinction efficiency, its subwavelength confinement and guiding. The photon energy of the resonances can be tuned in a broad range from the ultraviolet to the infrared through the control of the nanostructure composition, shape, size, and it is sensitive to the nature of the surrounding medium and possibly to external excitations (such as light, heat, voltage). Furthermore, their assembling into optical metamaterials opens unprecedented possibility for the engineering of light. These features have brought great promises for various applications, including energy conversion, passive and switchable elements for photonic circuitry, and nanoscale sensing.

So far, optical resonances in subwavelength nanostructures (i.e. with a size much smaller than the photon wavelength *in vacuo*) have been achieved mostly through the excitation of surface plasmons, i.e. collective oscillations of free charge carriers [1,2]. Resonance modes can be excited in plasmonic nanostructures several times smaller than the photon wavelength *in vacuo*. Initially, efforts were focused on noble metal (Ag and Au) nanostructures, which show an outstanding performance for achieving surface plasmon resonances tunable across the visible and near infrared regions. However, it is remarkable that plasmonic effects have been achieved using other materials, provided they behave optically as a "metal" in the spectral region of interest, with low enough optical losses. This requirement is met when the values of the material complex dielectric function ($\varepsilon = \varepsilon_1 + j\varepsilon_2$) show a negative real part ($\varepsilon_1 < 0$) and a small enough imaginary part $\varepsilon_2$. This important insight launched an intense search for "alternative plasmonic materials" [3], required especially for driving surface plasmon resonances in spectral regions beyond the visible and near-infrared. Al was shown to be an excellent candidate for achieving plasmonic effects in the ultraviolet [4]. Doped semiconductors have been shown to be suitable for achieving plasmonic effects tunable from the far to the mid infrared by controlling the dopant concentration [3] or by optical activation [5]. These works highlight the fact that the free charge carrier density N plays a crucial role on plasmonic effects, i.e. that the higher N the higher the photon energies at which plasmonic effects occur. The low free charge carrier density in moderately doped semiconductors limits plasmonic effects to the far infrared, whereas the high free charge carrier density of Al allows plasmonic effects in the ultraviolet. This trend is well described by the Drude dielectric function that accounts for the response of screened free charge carriers in a solid. Its real part $\varepsilon_1$ decreases with photon energy and turns from positive ($\varepsilon_1 > 0$) to negative ($\varepsilon_1 < 0$) at the *zero-crossing* photon energy $E_S$, which increases with the free charge carrier density/effective mass ratio $N^* = N/m^*$ [6]. The search for "alternative plasmonic materials" has also aimed at identifying materials that, although could not outperform the plasmonic response of noble metals in the visible and near infrared, could show some additional advantage. In this context, TiN has proven to be an alternative to Au in the visible, with comparable resonance quality factors and a better stability suitable for high temperature applications or non-linear optics [3, 7]. In addition, noble metals are not the best choices for switchable plasmonic devices due to their insensitivity to external excitations. This has triggered the search for plasmonic materials with switchable $\varepsilon$.

It must be underlined that the optical resonances in plasmonic nanostructures of any nature are inevitably linked with optical losses, because the excitation of free charge carriers involves photon absorption followed by thermal dissipation. These losses are beneficial for applications such as photothermal therapy, but are a drawback for others that require ultra-sharp resonances or a high optical throughput, such as ultrasensitive sensing or light guiding in photonic circuits.

In this context, a strong emphasis was made during the last years on optical resonances based on Mie modes [8-10]. Mie resonances are based on optical interferences, and thus they can be achieved using dielectric nanostructures consisting of transparent materials ($\varepsilon_2 = 0$). Therefore, in contrast with surface plasmon resonances, they do not require the excitation of free charge carriers and thus they have the advantage that they can be lossless. Mie resonances occur when the photon wavelength *in the nanostructure* is comparable with or smaller than the nanostructure size. Therefore, to achieve such resonances in nanostructures much smaller than the photon wavelength *in vacuo*, the constituting material must present a high value of the real part of its dielectric function ($\varepsilon_1 \gg 1$). Much attention has been paid to the semiconductor Si, which presents a high $\varepsilon_1$ ($\varepsilon_1 > 10$) in the visible and infrared and a low $\varepsilon_2$ (5 times lower than Ag at 2 eV, and $\varepsilon_2 = 0$ below 0.94 eV). This enables subwavelength Si nanostructures to support Mie resonances in the visible. Several modes are observed, the first one having a magnetic dipolar character, and the second one having an electric dipolar character. The existence of a magnetic dipolar mode in the visible (not achievable in plasmonic nanostructures) has several implications for the design of photonic circuitry. Especially, the tailoring of its interference with the electric dipolar mode allows tuning the directionality of the scattered light [11]. Due to the low cost and technological importance of Si, most of the experimental research about Mie resonances has involved Si nanostructures. However, there are plenty of other materials that in principle present suitable $\varepsilon_1$ and $\varepsilon_2$ values for achieving Mie resonances in subwavelength nanostructures [8, 12, 13] and could present advantageous properties over Si for certain specific applications. Ge has been pointed as one of these materials [12]. Ge, as Si, is a semiconductor and belongs to the p-block of the periodic table of elements. The similarity in electronic structure has appealed at evaluating the potential of other p-block materials, elements and compounds, for supporting Mie resonances [13].

Very recently, optical resonances have been reported at ultraviolet, visible and near infrared photon energies in subwavelength nanostructures made of semi-metal or topological insulators from the p-block [14-15]. These materials show negative $\varepsilon_1$ values in this spectral region, and their resonances were thus initially named "plasmonic". However, the cause of these resonances is not the excitation of free charge carriers, which is only relevant at much lower photon energies, but the excitation of strong interband transitions, possibly together with topological surface conducting states in some cases. This peculiar behavior is appealing for switchable plasmonics or energy conversion processes such as photocatalysis, and it is worth looking for similar effects in other materials, especially other semi-metals, semiconductors and topological insulators from the p-block.

In sum, there is nowadays a growing interest in semi-metals, semiconductors, and topological insulators from the p-block for the development of optically resonant nanostructures with subwavelength dimensions. In this paper, we evaluate the potential of such materials (single-element or compounds) for building subwavelength nanostructures supporting both plasmonic and Mie resonances. Especially, we show that this potential is tightly linked with the existence of strong interband transitions that drive both the plasmonic and Mie resonances, which are thus like the two sides of a single coin.

In section 2, we describe the concept behind this duality. In section 3, we illustrate it in the case of the semi-metal Bi and discuss which applications may arise based on the optically resonant character of subwavelength Bi nanostructures. In section 4, we review the optical properties of the single-element materials and of some selected binary, ternary and quaternary compounds from the p-block that show a semi-metal, semiconductor or topological insulator behavior. We discuss the potential of each of these materials in relation with plasmonic and Mie resonances. Furthermore, we report recent experimental demonstrations of these resonances in subwavelength nanostructures and discuss which applications could profit from them.

## 2. Concept: Strong interband transitions drive both plasmonic and Mie resonances

The complex dielectric function $\varepsilon = \varepsilon_1 + j\varepsilon_2$ of a material must obey the Kramers-Kronig relations. Therefore, a dielectric function dominated by an interband transition that can be depicted as an absorption band in the $\varepsilon_2$ spectrum will also present an anomalous jump in the $\varepsilon_1$ spectrum. If the oscillator strength of the interband transition is high, i.e. the peak value of $\varepsilon_2$ is high, as a result $\varepsilon_1$ will turn strongly negative at photon energies higher than the $\varepsilon_2$ peak energy, and will take high positive values at photon energies lower than the $\varepsilon_2$ peak energy [3, 16-19]. This phenomenon is illustrated in Fig. 1(a) using a dielectric function based on a single Kramers-Kronig consistent Lorentz oscillator with a peak $\varepsilon_2$ of 120 at a photon energy of ~ 1 eV, i.e. in the near infrared. Because of this high peak value, $\varepsilon_1$ takes strongly negative values above 1 eV, and high positive values ($\varepsilon_1 \sim 80$) below 0.6 eV. In addition, $\varepsilon_2$ tends towards 0 away from the $\varepsilon_2$ peak energy. Therefore, in subwavelength nanostructures with this dielectric function, the conditions for achieving plasmonic resonances ($\varepsilon_1 < 0$, small $\varepsilon_2$) are met in the ultraviolet-visible, while those for achieving Mie resonances ($\varepsilon_1 \gg 1$, small $\varepsilon_2$) are met in the mid infrared - far infrared.

To illustrate this behavior, Fig. 1(b) shows the simulated optical extinction efficiency of spherical nanoparticles with the dielectric function shown in Fig. 1(a), and different diameters D (25 nm and 800 nm). The 25 nm - nanoparticle shows a resonance at the boundary between the visible and ultraviolet, at a photon energy near 3 eV. This resonance, which originates from the negative $\varepsilon_1$ in this spectral region, has a plasmonic character. Note that the nanoparticle is much smaller than the resonance wavelength *in vacuo* (~ 400 nm). The 800 nm – nanoparticle shows several resonances in the mid infrared, at photon energies between 0.1 and 0.3 eV. These resonances, which originate from the positive $\varepsilon_1$ in this spectral region, have a Mie character. Because $\varepsilon_1 \gg 1$, the nanoparticle resonates at wavelengths *in vacuo* (between 12 μm and 4 μm) several times its diameter.

Summarizing, the strong Lorentz oscillator accounting for the interband transition is the origin of both plasmonic and Mie resonances in the nanoparticles. These resonances occur in well separated spectral regions and for nanoparticles with different characteristic sizes, which are however always much smaller than the resonance wavelength. Therefore, a material that presents interband transitions with a high oscillator strength is a good candidate for designing subwavelength nanostructures showing plasmonic and Mie resonances, at higher and lower photon energies than the interband transitions, respectively. The photon energies of these interband transitions will dictate the specific spectral regions where plasmonic and Mie resonances are allowed.

The amplitude and width of the interband transitions also play an important role. This is exemplified in Fig. 2(a) where the $\varepsilon_1$ of two Kramers-Kronig consistent Lorentz oscillators peaking at ~ 1 eV are plotted. One of the oscillators has a high amplitude and is sharp, and the other oscillator has a lower intensity and is broad, as shown in the inset of Fig. 2(a). The $\varepsilon_1$ of Ag [21] and Si [22] have been plotted for comparison. When the photon energy decreases below 1 eV, the $\varepsilon_1$ of both oscillators converge toward ~ 90, i.e. much higher than the $\varepsilon_1$ of Si whose interband transitions have a much lower oscillator strength. Therefore, the stronger the interband transitions, the higher $\varepsilon_1$ on their low photon energy side, and thus the smaller the size of nanoparticles that can support Mie resonances.

It can also be seen in Fig. 2(a) that the $\varepsilon_1$ of both the sharp and broad oscillators are comparable to those of Ag in the 1.5 – 4 eV photon energy range, whereas their $\varepsilon_2$ takes low values that can make their plasmonic figure of merit overcome that of Ag in a specific range of energies. This is apparent in Fig. 2(b) where we have plotted the surface plasmon quality factor $Q_{LSPR} = |\varepsilon_1|/\varepsilon_2$ [23, 24] that depicts the ratio between the local field around a plasmonic nanoparticle and the incident field. The dielectric function of the broad oscillator yields a better quality factor than Ag at photon energies above 3.4 eV, i.e. in the ultraviolet, while that of the sharp oscillator overcomes Ag already in the green-blue region of the visible.

In sum, the discussion above suggests that materials whose optical properties are dominated by strong interband transitions could show an enhanced plasmonic performance with respect to noble metals at visible frequencies, and could overcome Si as a platform for supporting Mie resonances in subwavelength nanostructures. This requires interband transitions with a high oscillator strength, with $\varepsilon_2$ peak values of 100 or even more. The question is therefore, can such values be found in any natural or artificial material?

## 3. Bismuth: A natural material with strong interband transitions

### 3.1 The dielectric function of bulk Bismuth

Bi is a semi-metal that early attracted the attention of scientists for the basic understanding of its peculiar physical properties, which include a very long charge carrier mean free path at room temperature, huge magnetoresistance, quantum confinement effects in nanofilms and nanowires with dimensions of several tens of nanometers. In this context, spectroscopy studies aimed at unveiling its bulk electronic band structure, and the dielectric function of Bi films or crystals was reported in different and limited spectral regions. These data, when plotted together as in [25], show a broad dispersion attributed to the variable sample quality. The presence of imperfections such as surface roughness or porosity, which were in many cases not characterized and not considered in the analysis of the experimental data, prohibited the access to the *true* bulk dielectric function of Bi.

Therefore, surprisingly there was until recently no report giving the dielectric function of bulk Bi in a broad spectral region. We filled this gap in the knowledge by performing spectroscopic ellipsometry measurements from the far infrared to the ultraviolet on high quality oriented Bi films, followed by an analysis involving the use of the transfer matrix formalism and Kramers-Kronig consistent oscillators plus a Drude function [25]. Our measurement was sensitive mostly to the in-plane response of the films. Therefore, we obtained the ordinary dielectric function of bulk Bi, which has a rhombohedral crystalline structure and thus presents a (likely slight) uniaxial optical anisotropy. The obtained dielectric function is shown in [25].

The broadband spectral profile of $\varepsilon_1$ and $\varepsilon_2$ is very like the very simple picture shown in Fig. 1(a). The $\varepsilon_2$ spectrum shows a strong peak (with values up to 120) near 0.8 eV that is attributed to interband transitions with a high oscillator strength. The $\varepsilon_1$ spectrum shows strongly negative values at higher photon energies and high positive values (up to 120) at lower photon energies. This suggests that subwavelength Bi nanostructures could support plasmonic resonances in the ultraviolet, visible and near infrared, and Mie resonances in the mid infrared – to – far infrared. Based on the analysis of the bulk dielectric function of Bi, we showed in [25] that it is totally driven by interband transitions from the ultraviolet to the mid infrared. Thus, we proposed that these interband transitions have the potential to *fully* induce the plasmonic resonances of Bi nanostructures, which we thus name hereafter "interband plasmonic" resonances. This is possible because the semi-metal character of Bi makes its N* ratio too low for free charge carriers to be excited in the ultraviolet to mid infrared. As seen in Fig. 3, the free charge carrier contribution to $\varepsilon$ starts to be sizeable only in the far infrared, where it makes $\varepsilon_1$ drop and $\varepsilon_2$ increase when moving toward lower photon energies. This sets a lower limit to the photon energy range in which Mie resonances could be observed in Bi nanostructures. At still lower photon energies in the far infrared, $\varepsilon_1$ becomes negative due to the contribution of free charge carriers thus making possible plasmonic properties. Such properties have been recently explored in Bi microstructures for applications in infrared sensing [26-27].

*3.2 Ultraviolet, visible and near infrared plasmonic resonances and mid – to – far infrared Mie resonances in subwavelength Bi nanostructures*

The first report we are aware of mentioning the term "surface plasmon resonance" for Bi nanostructures was published in 1995 by Park and coworkers [28], who postulated a plasmonic origin to the ultraviolet absorption band they observed in Bi nanoparticles in fused silica. Nevertheless, few attention was paid at the plasmonic potential of Bi prior to 2012 when we reported experimentally the tuning of the optical resonances of embedded Bi nanostructures from the ultraviolet to the near infrared as a function of their shape and size [14]. The tuning of these resonances is illustrated in Fig. 3(a), where the resonances are seen to occur at photon wavelengths *in vacuo* between 300 and 500 nm, for nanostructure sizes of a few tens of nanometers. Simulations of the optical extinction spectra of the nanostructures using their actual shape and size and the dielectric function of bulk Bi were in good qualitative agreement with the experimental data. This suggests that, at least in this spectral region and for the nanostructure sizes considered, the dielectric function of nanostructured Bi should note differ drastically from that of the bulk. Thus, it is very likely that the observed resonances have an interband plasmonic character.

This unconventional plasmonic character, which is based on the excitation of photo-generated electrons and holes, might be a key element in the photocatalytic behavior of Bi nanostructures that was demonstrated by Wang et al. and Dong et al. in 2014 [29, 30] and reported in many related works during the following years (see [31] and the references from 63 to 82 in [25]). At resonance, the optical absorption efficiency of Bi nanostructures can reach values of 5 [32]. Therefore, they are very efficient photon absorbers, and once the photons are absorbed they are directly converted in conduction electrons and valence holes that are able to activate chemical reactions at the nanostructure surface. Note that the same described concepts could be applied to develop wavelength- and polarization- selective photodetectors based on Bi nanostructures. Indeed, recently broadband high responsive photodetection in the visible and near infrared was reported at room temperature using flat Bi films [33]. In the photodetectors, photons are absorbed in the film volume, where they excite the Bi interband transitions (although the authors did not mention them). This generates electrons and holes, which then transfer to highly conducting surface states that allow their efficient migration to the electrodes. We envision that engineered nanostructured films, for instance consisting of an array of Bi nanostripes, could be considered for boosting "plasmonically" the optical absorption at desired wavelengths and polarizations, thus allowing to tailor the photonic response of the detector by structural design.

The plasmonic resonances of Bi nanostructures, together with the capability of Bi to undergo a solid-liquid phase transition at a lower temperature than most heavy elemental materials (270ºC), are useful for the design and fabrication of materials with a switchable absorption in a selected region of the optical spectrum [32, 34]. This switching is made reversible over many heating-cooling cycles provided the Bi nanostructures are embedded in a robust transparent matrix that remains unchanged during the cycles and act as a solid container for the Bi material. Upon melting (solidification), the dielectric function of the nanostructures changes from that of the solid (liquid) to that of the liquid (solid) Bi which are well differentiated in a broad spectral region: the dielectric function of liquid Bi is characteristic of a Drude metal with strong losses. In the ultraviolet – visible region, this contrast induces a shift in the plasmonic resonances of the nanostructures that translates into an optical absorption contrast at selected photon energies. We have proposed that these photon energies can be tuned using different matrices or by tuning the shape and size of the nanostructures [32, 34]. Especially, spherical Bi nanoparticles in a matrix with moderate refractive index and a size around 30 nm offer solutions for resonant switching in the near ultraviolet (photon wavelength in vacuo ~ 350 nm [32]).

Mie resonances in Bi nanostructures have been predicted theoretically in the mid to far infrared [25], nevertheless they have not observed experimentally. Fig. 3(b) shows the optical extinction spectrum in this spectral region of a 600 nm Bi spherical nanoparticle embedded in a transparent matrix with a common refractive index (1.65). The refractive index and extinction coefficient of Bi in the mid infrared remain near 9.5-10 and 1, respectively. Two Mie modes can be seen, at photon wavelengths *in vacuo* near 6 μm and 4.5 μm. We tentatively attribute them to the magnetic dipolar and electric dipolar Mie modes, respectively. By making an analogy with Si in the visible and near infrared, it can be expected that tuning the interference between these modes could result in switching the directionality of the light scattered by the nanoparticle. As presented in Fig. 4b, this effect is indeed predicted by Mie simulations that show scattering in opposite directions for different wavelengths of the incident photons: in the same direction as the incident beam (forward scattering) for a wavelength of 7.2 μm, and in the opposite direction (backward scattering) for a wavelength of 5.2 μm. It is remarkable that this effect takes place for a nanoparticle size of around one tenth of the wavelength thus opening the way to a markedly subwavelength manipulation of the propagation of mid infrared light using Bi nanostructures. Let us finally note that, although confinement effects were reported [35, 36] in Bi nanowires and shown to affect their dielectric function in the mid infrared, the structures studied were much smaller than those considered for supporting Mie resonances. We understand that the dielectric function of Bi nanostructures with diameters of several hundreds of nanometers should not differ much from that of the bulk material. However, a systematic investigation of the effect of confinement on the broadband dielectric function of Bi, with a broad variety of nanostructure shape and sizes, is missing and shall be an interesting topic for future works.

## 4. Exploring the p-block: In search of building blocks for subwavelength nanostructures with interband plasmonic and Mie resonances

### 4.1 Elemental materials

Other materials from the p-block share common features with Bi, i.e. dominant interband transitions that shape their bulk dielectric function. To visually identify them, we first show in Fig. 4 the bulk dielectric functions of the p-block single-element materials as distributed in the periodic table. We have selected for each element the values that we considered the most representative among those reported in the literature. Otherwise stated, we focused on the most frequent crystalline phase in ambient conditions and on works reporting the measurements on crystals or films whose properties are considered to represent well the bulk material. We discarded reporting on C (see extensive literature on graphite), O, N and the noble gas (not solid in ambient conditions) and the halogens (to reduce the number of binary compounds in the next section). Several of these materials present anisotropic optical properties. In the cases where the different components of the dielectric tensor were measured (for instance by measurements on single crystals), we represent only one of its components. In the case of uniaxial materials, we represent the ordinary dielectric function (optical response in the basal plane). In the case of biaxial materials, we have selected the dielectric function along the a-axis (optical response in one preferential direction of the basal plane). Note that except for elements of very strong technological interest such as Si and Ge or metals like Al, and similarly with Bi, it seems hard to find studies reporting the broadband bulk dielectric function of these materials from measurements on samples with an excellent and verifiable quality. Despite of this lack of fully accurate and reliable information, we assume that at least global qualitative trends can be derived from the available data. We have grouped them according to their behavior with a color code.

The dielectric functions of the metals Al, In, Tl, Sn (white tin) and Pb are, as expected, dominated by the Drude response of their free charge carriers. They therefore show a conventional plasmonic behavior and we will not consider them further. Therefore, these elements are represented in grey. In a similar way, the dielectrics B, P, S and Se, represented in light blue, are not of interest for the purpose of this review because they present small $\varepsilon_1$ values in the spectral region below their bandgap.

Si, Ge and As (amorphous arsenic) are semiconductors with strong interband transitions in the ultraviolet and visible. Due to their semiconductor character, these materials show no free charge carrier contribution at all from the ultraviolet to the far infrared. They show negative $\varepsilon_1$ values in the ultraviolet that are thus fully induced by the interband transitions, according to the Kramers-Kronig relations. The semiconductor Te shares common trends with these elements, however with a more symmetric $\varepsilon_2$ profile that makes its $\varepsilon_1$ turn negative already in the visible. Consequently, Si, Ge and As nanostructures are expected to display interband plasmonic effects in the ultraviolet, and Te nanostructures in both the ultraviolet and visible. Therefore, Si, Ge and As are represented in blue and Te, in green. Also for these elements, in accordance with the Kramers-Kronig relations, $\varepsilon_1$ is high at the low photon energy side of their interband transitions, where the optical absorption becomes negligible. This makes Si, Ge, As and Te very good candidates for building subwavelength nanostructures showing Mie resonances in the visible and/or infrared. Since their interband transitions are active in different spectral regions, the highest photon energy accessible to the Mie resonances differs from one material to another. Si shows a sharp drop in its interband absorption near 3 eV, and therefore is the most adequate candidate among the four for achieving Mie resonances in the visible.

Furthermore, due to the absence of free carrier absorption in the infrared for Si, Ge, As and Te, $\varepsilon_1$ does not drop thus preserving their capability for supporting Mie resonances also in this spectral region.

Bi and Sb are maybe the most interesting elements of the p-block, regarding their peculiar optical properties [53]. The properties of Sb are very similar to those of Bi. For these reasons, we have plotted them both in red. Sb is a semi-metal with strong interband transitions peaking in the short-wave infrared with a peak $\varepsilon_2$ value around 100. As in the case of Bi, there is no contribution of free charge carriers from the ultraviolet to the mid infrared, where the dielectric function is fully driven by the interband transitions. They make $\varepsilon_1$ take strongly negative values from the visible to the near infrared thus making Sb nanostructures prone to showing an interband plasmonic response in this region. As in the case of Bi, $\varepsilon_1$ reaches high values ($\varepsilon_1 \sim 80$) at the low photon energy side of the interband transitions. However, the N* of Sb is almost one order of magnitude higher than that of Bi. This makes the free charge carrier contribution become effective already in the mid infrared, where $\varepsilon_1$ starts to decrease. This limits the spectral range in which subwavelength Sb nanostructures can support Mie resonances.

Ga also presents strong interband transitions in the visible and near infrared. They drive almost fully its dielectric function in the ultraviolet-visible, where $\varepsilon_1$ becomes negative, however with a small contribution of free charge carriers. This contribution is made possible by the higher N* of Ga compared to that of Sb and Bi [53], Ga being a covalent metal (not a semi-metal). Therefore, Ga nanostructures have the potential to support *nearly* interband plasmonic resonances in the ultraviolet-visible. The spectral region where Ga shows a negative $\varepsilon_1$ is less extended toward low photon energies than in the case of Bi and Sb, thus we have plotted the dielectric function of Ga in orange. Note that that the free charge carrier contribution of Ga becomes strong already in the near infrared, where $\varepsilon_1$ drops to negative values. Therefore, it is unlikely that Ga nanostructures could support infrared Mie resonances.

*4.2 Binary compounds*

There are plenty of compounds formed by p-block elements only. In Figs. 5 and 6, we present selected bulk dielectric functions of some binary compounds found in the literature. We group them based on the columns to which belong the elements forming the compound: III-V and III-VI (Fig. 5), IV-VI and V-VI (Fig. 6).

III-V compounds show a behavior qualitatively comparable with Si or Ge. They are semiconductors showing strong or moderately strong interband transitions with two main peaks in the ultraviolet – visible. These interband transitions fully induce the negative values of $\varepsilon_1$ in the ultraviolet and high positive values of $\varepsilon_1$ at lower photon energies. Note that the spectral regions where $\varepsilon_1 < 0$ (interband plasmonic) and where $\varepsilon_1$ takes high positive values (Mie) depend on the nature of the compound, in relation with the position, amplitude, and width of the interband peaks. Qualitatively, it can be seen from Fig. 6 that both these spectral regions shift toward lower photon energies upon increasing the atomic mass of the components. Among the III-V compounds, AlP is especially remarkable because of its main absorption onset at 4 eV that makes it ideal for achieving Mie resonances in the near ultraviolet. III-VI chalcogenide compounds show similar properties with the III-V, however with weaker interband transitions and thus lower positive/less negative $\varepsilon_1$ values.

The IV-VI chalcogenide compounds (semiconductors, except SnTe and GeTe: semi-metals) and the V-VI chalcogenide compounds (topological insulators) show similar trends: strong interband transitions that *fully* induce negative $\varepsilon_1$ values at higher photon energies and high positive values at lower photon energies. The regions where $\varepsilon_1 < 0$ and $\varepsilon_1 > 0$ depend on the nature of the compound: they shift toward lower photon energies upon increasing the atomic mass of the components. However, the magnitude of this shift is much higher than with the III-V and III-VI compounds: the region where $\varepsilon_1 < 0$ (interband plasmonic) is limited to the ultraviolet for GeS, PbS, GeSe, and $Sb_2Se_3$, it extends from the ultraviolet to the visible for SnSe, PbSe and $Bi_2Se_3$, and from the ultraviolet to the near infrared for GeTe, SnTe, PbTe, $Sb_2Te_3$ and $Bi_2Te_3$. Furthermore, there are relevant differences in the spectral signature of the compounds. Those containing Te are dominated by one main absorption band, asymmetric and rather sharp with peak $\varepsilon_2$ values as high as 50, that makes $\varepsilon_1$ reach values near 50 at lower photon energies. Note that the spectral region of high $\varepsilon_1$ should be quite narrow for GeTe due to its semi-metal character and its relevant free charge carrier contribution already in the mid infrared.

### 4.3 Performance of the single-element and binary compounds for interband plasmonic and Mie resonances

The semiconductors and topological insulators described above show similar characteristics in the sense that their bulk dielectric function is dominated by interband transitions with high oscillator strength, which *fully* induce negative and high positive $\varepsilon_1$ values on their high/low photon energy sides, respectively. Semi-metals display a comparable behavior, except in the far infrared (sometimes also in the mid infrared) where free charge carriers make $\varepsilon_1$ values drop and $\varepsilon_2$ values increase. Therefore, subwavelength nanostructures built from these materials are expected to support interband plasmonic and Mie resonances, in spectral regions that can be selected upon adequate material choice. In this section, we present for each material the ranges where interband plasmonic and Mie resonances can be supported, and evaluate the performance of these resonances using simple and well known indicators.

For interband plasmonic resonances, we represent the figure of merit $Q_{LSPR} = |\varepsilon_1|/\varepsilon_2$ as a function of photon energy for the different materials (Fig. 7(a)). This figure of merit, in addition to representing the intensity of the near field of a nanostructure, provides qualitative information about the sharpness of the interband plasmonic resonances achievable with the different materials (a high/low $Q_{LSPR}$ stands for sharp/broad resonances). The $Q_{LSPR}$ obtained for the considered materials reaches values no higher than 3. This implies that the interband plasmonic resonances of the nanostructures show a low near field enhancement (as shown by the calculations by McMahon et al. in the case of Bi and Ga [63]), and a broad spectral width (as observed in the case of Bi nanostructures [14, 32]). This discards the use of these materials for applications that require strong field confinement or sharp resonances, such as surface enhanced spectroscopy or ultrasensitive sensing. However, as seen in the case of Bi [14, 32], a $Q_{LSPR}$ near 1 is sufficient to achieve well defined resonances with a high extinction efficiency. This is especially what is needed for the light harvesting in energy conversion processes such as photocatalysis and for switchable plasmonics.

Si, Ge, As and the III-V and III-VI semiconductors can support interband plasmonic resonances in the ultraviolet only. Ga, Te and the IV-VI and V-VI compounds show a broader photon energy range of operation, with the broadest bandwidth for Ga and the tellurides, and especially $Bi_2Te_3$ that covers the whole ultraviolet - visible. A still broader bandwidth is shown by Bi and Sb, which can support resonances up to the near infrared region. It is worth noting that all these materials show a better plasmonic figure of merit than Ag only in the ultraviolet region, but overcome Au already in the blue-green region of the spectrum.

For Mie resonances, we represent the value of $\varepsilon_1$ at the photon energy of the main onset of absorption as a function of that energy (Fig. 7(b)). At this photon energy, $\varepsilon_2$ is already very low and $\varepsilon_1$ is already high, thus allowing Mie resonances in subwavelength nanostructures. In the case of semi-metals, we chose the value of $\varepsilon_1$ at the photon energy where $\varepsilon_2$ is minimum. In this plot, we get a visual representation of the highest photon energy at which Mie resonances can be obtained for a given material (the onset photon energy), and about the minimum size/photon wavelength *in vacuo* ratio required for a nanostructure to support Mie resonances just below this photon energy. The higher the $\varepsilon_1$ value, the smaller the characteristic size of the resonant nanostructures. Therefore, if one aims at designing Mie resonant nanostructures as small as possible compared with the photon wavelength *in vacuo*, then a as high as possible $\varepsilon_1$ is required.

Fig. 7(b) shows similar $\varepsilon_1$ values in the 10-20 range for almost all the materials. The compound AlP shows a similar $\varepsilon_1$ with Si, but a higher onset photon energy, in the near ultraviolet. Therefore, it is suitable for achieving Mie resonances in the near ultraviolet, a region that cannot be covered by Si nanostructures. The resonant AlP nanostructures can show a size/photon wavelength *in vacuo* ratio like that of Si nanostructures for visible Mie resonances. The highest $\varepsilon_1$ values are found for the materials with an onset photon energy in the near infrared and short-wave infrared: tellurides, Bi and Sb, with values of up to 100 for Bi (however at the expense of a non-zero $\varepsilon_2$). Among them, Bi, $Bi_2Te_3$, $Sb_2Te_3$ and PbTe are promising for achieving Mie resonances in the mid infrared (photon wavelengths in vacuo between 4 and 10 μm) with nanostructures (smaller than 1 μm) because of their remarkably high $\varepsilon_1$ in this region, which is permitted by the absence of free charge carrier contribution.

*4.4 Ternary and quaternary compounds*

We will not review in detail the optical properties of all the ternary and quaternary chalcogenide compounds from the p-block, because of the extremely large number of combinations. We will just stress the importance of two classes of compounds: the GSTs (germanium antimony tellurium) and the BSTs & BSTSs (bismuth selenium tellurium & bismuth selenium tellurium antimony).

GSTs are well known phase transition materials that can be driven from an amorphous to crystalline state *et vice versa* by conventional heating/cooling or using ultrashort laser pulses. Consequently, they have been extensively used for data storage. Shportko et al. [64] reported the dielectric function of GSTs with several compositions in the amorphous and crystalline state. The crystalline phases show a semiconductor character with strong interband transitions peaking in the near infrared, which induce negative $\varepsilon_1$ in the ultraviolet – visible and high $\varepsilon_1$ from the near/short-wave infrared down to the far infrared. The interband plasmonic and Mie performances using crystalline GSTs are very like those of telluride binary compounds shown in Fig. 7. The performances of the corresponding amorphous phases are much lower, in relation with the broader and less intense interband transitions. The drastic contrast between the optical properties of the amorphous and crystalline GSTs is thus ideal for switchable photonics.

BSTs and BSTSs belong to the last generation of topological insulators which have attracted the attention of the scientific community in relation with their outstanding surface transport properties. The experimentally determined dielectric function of the $Bi_{1.5}Sb_{0.5}Te_{1.8}Se_{1.2}$ was reported by Ou et al. [15], and Dubrovkin et al. [65]. Yin et al. [66] reported quantum simulated dielectric functions for several ternary and quaternary compounds: $Bi_2Se_2Te$, $Bi_2SeTe_2$, $BiSbTeSe_2$, $BiSbTe_2Se$. According to these experimental data and simulations, BSTs and BSTSs show interband plasmonic and Mie performances comparable with those of the telluride binary compounds in Fig. 8. However, the dielectric functions of most of the BSTs and BSTSs remain to be experimentally determined, and such study might be relevant to identify the performance of those among the many different achievable compositions [67].

## 4.5 Observation of interband plasmonic and Mie resonances in subwavelength p-block nanostructures and applications

### 4.5.1. Bi, Sb and Ga

So far, the resonance properties of Bi and Ga nanostructures have been studied experimentally only in the ultraviolet, visible and near infrared, in relation with plasmonics. We could find no experimental report about Sb nanostructures nor about Mie resonances in Bi and Ga nanostructures.

In addition to the references cited in section 3 about the ultraviolet – to – near infrared interband plasmonic properties of Bi nanostructures [14, 25, 28-32, 34], we would like to mention other works where such resonances seem to have been reported using optical spectroscopy techniques [68-70]. In particular, Sivaramakrishnan et al. reported the optical nonlinear response of Bi nanorods [70], however without discussing the underlying mechanism. It would be interesting to investigate this mechanism that might lead to very different effects compared with metal nanostructures due to the interband – based optical absorption process [71].

The plasmonic properties of Ga nanoparticles with sizes of tens of nm were reported by Knight et al. [72] using cathodoluminescence microscopy. Their measurement showed surface resonances in the visible region. Yang et al. reported ultraviolet resonances in Ga nanoparticles with sizes around 20-30 nm using spectroscopic ellipsometry [73]. Because of the relatively low melting temperature of Ga (around 30ºC for bulk Ga, and even much lower for nanoparticles), it is thus likely that the reported resonances have a "traditional" plasmonic origin related with the lossy Drude metal character of liquid Ga [72,73]. Due to this relatively low melting temperature and the significant contrast between the dielectric functions of its liquid and solid states, Ga attracted early the attention for ultrafast optically switchable nanophotonics [74-76]. Although efficient switching was reported for the propagating plasmons in Ga gratings upon melting-solidification around room temperature [77], we were not able to find any report about switching using Ga nanostructures supporting localized plasmon resonances. One of the reason to this lack of reports might be the strong lowering of the melting point of Ga due to confinement in nanostructures, thus requiring operations with active cooling that makes the switching process less practical than near room temperature. In this aspect, Bi might be a good alternative to Ga for ultrafast optically switchable nanophotonics because its melting point can be maintained above room temperature even in very few nm nanostructures [34, 78] thus requiring no active cooling.

### 4.5.2. Si, Ge, IV-VI and V-VI semiconductors

Numerous works have reported the properties of Mie resonances in subwavelength Si nanostructures [8-11] and the design of lossless metamaterials and metasurfaces. Some works have shown the possibility to tune or switch Mie resonances through the control of the dielectric properties of the Si material. Lewi et al. [79] demonstrated the tuning of infrared Mie resonances in Si resonators by doping-induced introduction of free carriers. Makarov et al. [80] and Baranov et al. [81] have reported switchable Mie resonances by pulsed laser excitation at photon energies above the bandgap of Si.

There has been an early interest in lead chalcogenide (PbS, PbSe, PbTe) nanostructures especially due to their strong confinement effects that allow a size-tunable bandgap across the infrared [82]. As discussed above, these materials are also interesting for supporting Mie resonances in much bigger nanostructures. PbTe is particularly relevant due to its higher $\varepsilon_1$ in the infrared (~40) than most of the semiconductors. Very recently, Lewi et al. demonstrated high quality factor mid infrared Mie resonances in PbTe subwavelength resonators with a size up to 8 times smaller than the photon wavelength *in vacuo* [83]. Due to the very high thermo-optical coefficient of PbTe compared with other semiconductors (e.g. Si, Ge or the III-V or III-VI compounds), they proposed that PbTe enables the fabrication of mid infrared ultra-narrow active notch filters and metasurface phase shifters based on a thermal modulation of a few ºC.

Assuming that the thermo-optical coefficient of a semiconductor at a selected photon energy increases with the slope of $\varepsilon_1$ vs photon energy (which is the case for Si), we suggest that an even more sensitive thermal modulation could be achieved using materials with stronger interband transitions than PbTe, at the low energy side of these transitions where $\varepsilon_1$ decreases very fast with photon energy.

Other technologically interesting semiconductors are the amorphous arsenide chalcogenides [84-86], which present moderately strong interband transitions in the visible and therefore moderately high $\varepsilon_1$ (~8-9) in the infrared region. Although these values do not compete with those exhibited by PbTe, the amorphous nature of the arsenide chalcogenides together with their very low achievable infrared optical losses ($\varepsilon_2 \rightarrow 0$) have made possible the fabrication of high-quality factor (~ $10^5$) whispering gallery mode microresonators on various platforms including flexible substrates. This is appealing for metamaterial-based optically-monitored sensing in the mid infrared which is nowadays an emerging topic due to the commercial availability of tunable lasers in this spectral region.

*4.5.3. GSTs*

The contrast between the dielectric functions of crystalline and amorphous GSTs made them very attractive for switchable plasmonics [87-88] enabling applications such as active optical filtering for mid infrared multispectral imaging. However, the GST was mostly used as a tunable matrix used for shifting the resonances of conventional plasmonic nanostructures (made for instance of Al). It is more recently [89] that GSTs have been used as the building block of optical resonators. Based on this approach, Karvounis et al. demonstrated all-dielectric phase-change switchable metasurfaces [89]. In their work, the GST material is structured in an array of parallel nanowires that exhibit grating resonances. Upon crystalline-to-amorphous phase transition, the grating resonances shift thus enabling a strong change in the metasurface transmittance and reflectance at selected photon energies. The crystalline phase of their GST exhibits strong interband transitions peaking at a photon wavelength *in vacuo* lower than 1 μm. This enables a high positive $\varepsilon_1$ and a low $\varepsilon_2$ already in the near infrared region. Upon transition to the amorphous phase, the oscillator strength of the transitions drops, thus inducing a big decrease in $\varepsilon_1$ that enables the switchable optical response of the metasurface. This appeals at designing of a broad variety of Mie-resonant GST nanostructures that will be used as building blocks of all-dielectric, low-loss metasurfaces with switchable anomalous reflection and optical phase shifts [90].

*4.5.4. V-VI topological insulators*

The optical and resonant properties of nanostructures based on V-VI topological insulators are since recently focusing the attention [15, 65, 91-96]. Resonances have been reported in the ultraviolet – visible and near infrared in nanostructures made of the binary ($Bi_2Te_3$ [91, 92] and $Bi_2Se_3$ [93]), ternary ($Bi_2Se_{1.4}Te_{1.6}$ and $Bi_2Se_{0.6}Te_{2.4}$ [92]), or quaternary ($Bi_{1.5}Sb_{0.5}Te_{1.8}Se_{1.2}$ [15, 65]) compounds. By means of optical reflectance spectroscopy and cathodoluminescence, Ou et al. [15] observed resonances on an array of nanoslits (typical dimensions ~ 200 nm) at the surface of a $Bi_{1.5}Sb_{0.5}Te_{1.8}Se_{1.2}$ bulk, and concluded at their plasmonic origin. Zhao. et al. did optical absorbance [92] and EELS [91] measurements on $Bi_2Te_3$ nanoplates with lateral dimensions and height of hundreds and tens of nm, respectively. They also concluded at a plasmonic origin of the resonances.

Zhao et al. [92] also demonstrated an active switching of these resonances upon crystalline to amorphous phase transition and discussed their potential for enhancing the performance of polymer solar cells. Guozhi et al. [93] proposed that $Bi_2Se_3$ nanostructures are suitable for enhanced photothermal conversion. The different authors focused on understanding the origin of the resonances in these nanostructures, which could arise from bulk interband plasmonic effects and also involve topological surface conducting states. In quaternary nanostructures, Ou et al. and Dubrovkin et al. [15, 65] proposed that the resonances are driven by both bulk and topological surface channels. They showed that resonances can be excited at photon energies below the zero-crossing energy of the bulk $\varepsilon_1$ (in the red region of the spectrum), and attributed them to surface plasmons driven by the topological surface conducting states whose $\varepsilon_1$ is negative in this region [65]. This finding suggests that very low loss plasmonic properties could be achieved if one could get rid of the bulk losses. This also shows great promises for applications in electrically switchable plasmonics, in relation with the recent demonstration of the gate tuning of the interband transitions in topological insulators [97] that opens the way to the tuning of the balance between bulk and topological surface channels.

Some interesting works pointed out that p-block topological insulators such as $Bi_2Se_3$ and $Bi_2Te_3$ should present a strong bulk optical anisotropy in relation with their lamellar Van der Waals structure [98, 99], which will impact their interband plasmonic properties. Such optical anisotropy can be found in other materials such as cuprates, ruthenates, magnesium diboride or graphite [100, 101] and can be even suitable for achieving hyperbolic optical properties in specific spectral regions. However, in most of the works reporting the bulk dielectric function of p-block topological insulators [61, 62, 102-105], anisotropy could not be properly resolved or was totally disregarded in the data analysis. Therefore, it is not clear yet whether the published data yield the in-plane dielectric function of the materials or a "pseudo" dielectric function mixing the in-plane and vertical dielectric function of the optically uniaxial material. In this context, Esslinger et al. [98] reported anisotropic dielectric functions of cleaved $Bi_2Se_3$ and $Bi_2Te_3$ single crystals. They extracted these values performing ellipsometry only on one crystal orientation at different angles, and in the data analysis they included optical anisotropy in their model. The resulting dielectric functions show strong interband transitions which show maxima at clearly different photon energies perpendicular to and along the optical axis of the crystals. Consequently, the corresponding negative $\varepsilon_1$ values occur in different spectral regions thus inducing the so-called optical hyperbolism. It is interesting to note that the resulting $\varepsilon_2$ values in the regions of negative $\varepsilon_1$ values are much smaller than those obtained by the other authors who disregarded anisotropy, thus suggesting a much higher interband plasmonic figure of merit than expected for these materials. Furthermore, theoretical predictions showed that these hyperbolic properties based on strong interband transitions can be exploited for applications such as hyper resolution imaging and subwavelength lithography, but they are still to be developed experimentally. Finally, a recent report by Talebi et al. [99] argues in favor of the existence of a strong anisotropy in $Bi_2Se_3$. The authors explored by EELS the optical modes in $Bi_2Se_3$ waveguides and resonators and attributed their spatial and spectral profiles to low loss Dyakonov waves and Dyakonov plasmons allowed by the hyperbolic optical response of $Bi_2Se_3$.

## 5. Conclusions and Perspectives

Semi-metals, semiconductors and topological insulators from the p-block, under the form of single-elements or compounds, have motivated and attracted the interest of scientists during the past decades for applications as varied as thermoelectricity, microelectronics and optoelectronics. It is only recently that these materials became appealing for the fabrication of subwavelength resonant nanostructures suitable for nanophotonics, plasmonics and even metamaterial engineering. In this context, we have analyzed in this manuscript the potential of all the elements of the p-block and some relevant compounds for the design of nanostructures supporting plasmonic and Mie resonances, by analyzing as a the first step their bulk optical properties in a broad spectral range from the infrared to the ultraviolet. From this analysis, it is evidenced that all these materials show an optical response dominated by strong interband transitions, which make the real part of the material's dielectric function ($\varepsilon = \varepsilon_1 + j\varepsilon_2$) take negative ($\varepsilon_1 < 0$) and high positive ($\varepsilon_1 \gg 1$) values at their high and low photon energy sides, respectively.

The study shows that both semi-metal and semiconductor subwavelength nanostructures are suitable for supporting so-called "interband plasmonic" resonances in the range of photon energies where $\varepsilon_1 < 0$. This range depends on the spectral position of the interband transitions and thus on the nature of the material: in the ultraviolet for III-IV and III-V compounds, in the ultraviolet to near infrared for elemental semi-metals Bi and Sb. Such resonances were reported experimentally for the first time in the ultraviolet to near infrared on Bi nanostructures [14]. Plasmonic resonances were also reported in the ultraviolet to near infrared on topological insulator nanostructures, for which they may involve topological surface conducting states in addition to the "bulk" interband plasmonic contribution [15, 65, 91]. Owing to their reported low quality factor, interband plasmonic resonances should lead to applications that require neither strong field enhancements nor very narrow resonances but exploit the specific physico-chemical properties of the p-block materials. Especially, in contrast with metals, whose plasmon resonances are due to the excitation of free charge carriers, interband plasmonic resonances are linked with the generation of photogenerated electron-hole pairs and are thus especially appealing for energy conversion processes such as photocatalysis. We infer that, in a reciprocal way, interband plasmonic resonances could be tuned electrically.

The spectral region in which subwavelength nanostructures can support Mie resonances (where $\varepsilon_1 \gg 1$ and $\varepsilon_2$ is small) and the characteristic structure sizes depend on: the spectral position of the interband transitions, their strength and the possible contribution of free charge carriers to $\varepsilon$. These features are dictated by the nature of the material constituting the nanostructure. AlP nanostructures are promising for achieving Mie resonances in the ultraviolet, whereas PbTe nanostructures and Bi nanostructures have been shown to be an excellent candidate for supporting mid infrared Mie resonances. Because they are fundamentally linked with low losses, Mie subwavelength resonant nanostructures are very attractive for the design of high optical throughput metamaterials or metasurfaces.

Because both Mie and plasmon resonances in p-block subwavelength nanostructures are driven by interband transitions, therefore they are very sensitive to the electronic band structure of the material. This property is very appealing for switchable plasmonics and nanophotonics, where Mie or interband plasmonic resonances would be tuned by changing the electronic configuration of the material. Phase transition-induced switching in GSTs or semi-metals is indeed based on this concept, and we foresee that many other switching approaches (involving the tuning of the occupancy of electronic states using light or voltage) will be demonstrated in the near future.

In sum, we have identified a handful of p-block "interband materials" with an interesting potential for the fabrication of non-conventional subwavelength resonant nanostructures. Relevant properties of nanostructures made of some of these materials have been recently reported. However, much work remains to be done for the basic scientific understanding of their properties, as well as for demonstrating suitable applications. Especially, it is necessary to explore the optical, optoelectronic and switching response of a broad variety of nanostructure shapes, sizes and of metamaterials based on them, and the feasibility of device development for the interband materials that have already shown promising properties.

Finally, it is very likely that a much broader range of interband materials with enhanced properties will be discovered in the near future. To name a few, many ternary and quaternary p-block compounds can be readily synthesized and their composition can be accurately controlled. How does this control affect their optical properties? Moreover, it is also very probable that similar optical properties can be found in compounds involving non p-block elements, or even without any p-block element at all. We can cite here very recent reports by Yan et al. and Braic et al. [106, 107] who demonstrated plasmonic effects at visible photon energies in substoechiometric titanium oxide and titanium oxynitrides, respectively. In the first case, an interband plasmonic origin was claimed for the resonances. In the second case, it was proposed that the oxynitride consists of a mixture of two phases that mimics at the macroscopic scale the visible optical response of an interband plasmonic material, in an analogous way to an earlier report from De Zuani et al. [108] that showed huge effective $\varepsilon_1$ values (up to 1000!) in nearly closed Au films. From a fundamental point of view there are two questions that will drive this research further: Is it possible to make materials whose interband plasmonic resonances overcome the performance of noble metals in the visible? How high can be made $\varepsilon_1$ below interband transitions?

It is interesting to note that, so far, nanophotonics and plasmonics have grown as a byproduct of wave optics mainly applied to noble metal, silicon and common dielectric structures. Therefore, it has remained rather disconnected from other fields of science such as solid state physics, electronics, quantum physics, crystallography, although the recent quest for alternative plasmonic materials showed the way to bridge this gap. Therefore, in a comparable way with the growth of electronics since the 50s, we expect that the quest for enhanced interband materials will be a new paradigm in nanophotonics and plasmonics by strongly extending their material platforms and boosting their interaction with many other fields of science and engineering.

# References

[1] U. Kreibig, and M. Vollmer, *Optical properties of metal clusters* (Springer, 1995).
[2] S.A. Maier, *Plasmonics: fundamentals and applications* (Springer, 2007).
[3] G. V. Naik, V. M. Shalaev, and A. Boltasseva, "Alternative plasmonic materials: beyond gold and silver," Adv. Mater. **25**(24), 3264–3294 (2013)
[4] M. B. Ross and G. C. Schatz, "Aluminium and indium plasmonic nanoantennas in the ultraviolet," J. Phys. Chem. C **118**(23), 12506–12514 (2014).
[5] M.P. Fischer, C. Schmidt, E. Sakat, J. Stock, A. Samarelli, J. Frigerio, M. Ortolani, D.J. Paul, G. Isella, A. Leitenstorfer, P. Biagioni, and D. Brida, "Optical activation of germanium plasmonic antennas in the mid-infrared," Phys. Rev. Lett. **117**, 047401 (2016).
[6] J. M. Luther, P. K. Jain, T. Ewers, and A. P. Alivisatos, "Localized surface plasmon resonances arising from free carriers in doped quantum dots," Nat. Mater. **10**(5), 361–366 (2011).
[7] N. Kinsey, A.A. Syed. D. Courtwright, C. DeVault, C.E. Bonner, V.I. Gavrilenko, V.M. Shalaev, D.J. Hagan, E.W. Van Stryland, and A. Boltasseva, "Effective third-order nonlinearities in metallic refractory titanium nitride thin films," Opt. Mater. Expr. **5**(11), 2395-2403 (2015).
[8] Y. Kivshar and A.E. Miroshnichenko, "Meta-optics with Mie resonances," Opt. Photon. News (January 2017)
[9] A.I. Kuznetsov, A.E. Miroshnichenko, Y.H. Fu, J. Zhang, and B. Luk'yanchuk, "Magnetic light," Sci. Rep. **2**, 492 (2012).
[10] A.B. Evlyukhin, S.M. Novikov, U. Zywietz, R.L. Eriksen, C. Reinhardt, S.I. Bozhevolny, and B.N. Chichkov, "Demonstration of the magnetic dipole resonances of dielectric nanospheres in the visible region", Nano Lett. **12**(7), 3749-3755 (2012).
[11] Y.H. Fu, A.I. Kuznetsov, A.E. Miroshnichenko, Y.F. Yu, and B. Luk'yanchuk, "Directional visible light scattering by silicon nanoparticles," Nat. Commun. 4, 1527 (2013).
[12] R. Gómez-Medina, B. Garcia-Cámara, I. Suárez-Lacalle, F. González, F. Moreno, M. Nieto-Vesperinas, and J.J. Sáenz, "Electric and magnetic dipolar response of germanium nanospheres: interference effects, scattering anisotropy and optical forces," Jour. Nanophoton. 5, 053512 (2011).
[13] B. García-Cámara, R. Gómez-Medina, J.J. Sáenz, and B. Sepulveda, "Sensing with magnetic dipolar resonances in semiconductor nanospheres," Opt. Expr. **21**(20), 23007-23020 (2013).
[14] J. Toudert, R. Serna, and M. Jiménez de Castro, "Exploring the optical potential of nano-Bismuth: Tunable surface plasmon resonances in the near ultraviolet to near infrared range," J. Phys. Chem. C **116**(38), 20530–20539 (2012).
[15] J. Y. Ou, J. K. So, G. Adamo, A. Sulaev, L. Wang, and N. I. Zheludev, "Ultraviolet and visible range plasmonics in the topological insulator $Bi_{1.5}Sb_{0.5}Te_{1.8}Se_{1.2}$," Nat. Commun. **5**, 5139 (2014).
[16] T. Pakizeh, "Optical absorption of plasmonic nanoparticles in presence of a local interband transition," J. Phys. Chem. C **115**, 21826-21831 (2011).
[17] T. Pakizeh, "Optical absorption of nanoparticles described by an electronic local interband transition," J. Opt. **15**, 025001 (2013).
[18] Z. Pirzadeh, T. Pakizeh, V. Miljkovic, C. Langhammer, and A. Dmitriev, "Plasmon-interband coupling in nickel nanoantennas," ACS Photonics **1**, 158-162 (2014).
[19] J. Toudert, "Spectroscopic ellipsometry for active nano- and meta- materials," Nanotechnol. Rev. **3**(3), 223–245 (2014).
[20] MiePlot software – www.philiplaven.com/mieplot.htm
[21] E.D. Palik, *Handbook of optical properties of solids* (Academic Press, 2008).
[22] P. Lautenschlager, M. Garriga, L. Vina, and M. Cardona, "Temperature dependence of the dielectric function and interband critical points in silicon," Phys. Rev. B **36**, 4821 (1987).
[23] Y. Jiang, S. Pillai, and M. Green, "Realistic silver optical constants for plasmonics," Sci. Rep. **6**, 30605 (2016).
[24] P.R. West, S. Ishii, G.V. Naik, N.K. Emani, V.M. Shalaev, and A. Boltasseva, "Searching for better plasmonic materials," Laser & Photon. Rev. **4**(6), 795-808 (2010).
[25] J. Toudert, R. Serna, I. Camps, J. Wojcik, P. Mascher, E. Rebollar and T.A. Ezquerra, "Unveiling the far infrared-to-ultraviolet optical properties of bismuth for applications in plasmonics and nanophotonics," J. Phys. Chem. C **121**(6), 3511-3521 (2017).
[26] F. Khalilzadeh-Rezaie, C.W. Smith, J. Nath, N. Nader, M. Shahzad, J.W. Cleary, I. Avrutsky, and R.E. Peale, "Infrared surface polaritons on bismuth," J. Nanophotonics **9**, 093792 (2015).
[27] J.W. Cleary, G. Mehdi, R.E. Peale, W.R. Buchwald, O. Edwards, and I. Oladeji, "Infrared surface plasmon resonance biosensor," Proc. SPIE **767306** (2010).
[28] S.Y. Park, R.A. Weeks, and R.A. Zuhr, "Optical absorption by colloidal precipitates in bismuth implanted fused silica: annealing behavior," Appl. Phys. Lett. **69**, 3175 (1996).
[29] Z. Wang, C. Jiang, R. Huang, H. Peng, and X. Tang, "Investigation of the optical and photocatalytic properties of bismuth nanospheres prepared by a facile thermolysis method," J. Phys. Chem. C 118, 1155 (2014).
[30] F. Dong, T. Xiong, Y. Sun, Z. Zhao, Y. Zhoy, X. Feng, and Z. Wu, "A semimetal bismuth element as a direct plasmonic photocatalyst," Chem. Commun. 50, 10386-10389 (2014).
[31] W. Fan, C. Li, H. Bai, Y. Zhao, B. Luo, Y. Li, Y. Ge, W. Shi, and H. Li, "An in situ photoelectroreduction approach to fabricate Bi/BiOCl heterostructure photocathodes: understanding the role of Bi metal for solar water splitting," J. Mater. Chem. A 5, 4894-4903 (2017).

[32] A. Cuadrado, J. Toudert, and R. Serna, "Polaritonic-to-plasmonic transition in bismuth nanospheres for high-contrast ultraviolet meta-filters," IEEE Photonics J. **8**, 1 (2016).
[33] J.D. Yao, J.M. Shao, and G.W. Yang, "Ultra-broadband and high-responsive photodetectors based on bismuth film at room temperature," Sci. Rep. **5**, 12320 (2015).
[34] M. Jiménez de Castro, F. Cabello, J. Toudert, R. Serna, and E. Haro-Poniatowski, "Potential of bismuth nanoparticles embedded in a glass matrix for spectral-selective thermos-optical devices", Appl. Phys. Lett. **105**, 113102 (2014).
[35] M. R. Black, Y. M. Lin, S. B. Cronin, O. Rabin, and M. S. Dresselhaus, "Infrared absorption in bismuth nanowires resulting from quantum confinement," Phys. Rev. B **65**(19), 195417 (2002).
[36] M. R. Black, P. L. Hagelstein, S. B. Cronin, Y. M. Lin, and M. S. Dresselhaus, "Optical absorption from an indirect transition in bismuth nanowires," Phys. Rev. B **68**(23), 235417 (2003).
[37] N. Morita, and A. Yamamoto, "Optical and electrical properties of boron," Jap. J. Appl. Phys. 14(6), 825-831 (1975).
[38] O. Hunderi, and R. Ryberg, "Band structure and optical properties of gallium," J. Phys. F: Met. Phys. **4**(11), 2084 (1974).
[39] R.Y. Koyama, N.V. Smith, and W.E. Spicer, "Optical properties of indium," Phys. Rev. B **8**, 2426 (1973).
[40] G. Jezequel, J.C. Lemonnier, and J. Thomas, "Optical properties of gallium films between 2 and 15 eV," J. Phys. F: Met. Phys. **7**(8), 1613 (1977).
[41] G. Jezequel, J. Thomas, and I. Pollini, "Optical properties of thallium films studied with synchrotron radiation," Phys. Rev. B **37**(15), 8639 (1988).
[42] Si.mat datafile, WVASE32 software (J.A. Woollam Co.).
[43] UNL database, WVASE32 software (J.A. Woollam Co.).
[44] H.G. Liljenvall, A.G. Mathewson, and H.P. Myers, "The optical properties of lead in the energy range 0.6-6 eV," Phil. Mag. **22**(176), 243-253 (1970).
[45] N. Mao, J. Tang, L. Xie, J. Wu, B. Han, J. Lin, S. Deng, W. Ji, H. Xu, K. Liu, L. Tong, and J. Zhang, "Optical anisotropy of black phosphorus in the visible regime," J. Am. Chem. Soc. **138**, 300-305 (2016).
[46] A. Vaidyanathan, Y.F. Tsay, and S.S. Mitra, "Electronic structure and optical properties of rhomboedral arsenic and their variations in the internal displacement parameter," Phys. Stat. Sol. (b) **92**, 73-81 (1979).
[47] T.J. Fox, R. P. Howson, and D.C. Emmony, "Optical properties of thin films of antimony," J. Phys. D: Appl. Phys. **7**, 1864-1872 (1974).
[48] J.C. Lemonnier, J. Thomas, and S. Robin, "Optical properties and electronic structure of antimony in the energy range 2.5 – 14.5 eV," J. Phys. C: Solid State Phys. **6**, 3205 (1973).
[49] R. Sasson, R. Wright, E.T. Arakawa, B.N. Khare, and C. Sagan, "Optical properties of solid and liquid sulfur at visible and infrared wavelengths," Icarus **64**, 368-374 (1985).
[50] S. Tutihasi, and I. Chen, "Optical properties and band structure of trigonal selenium," Phys. Rev. **158**(3), 623 – 630 (1967).
[51] P. Bammes, R. Klucker, E.E. Koch, T. Tuomi, "Anistropy of the dielectric constants of trigonal selenium and tellurium between 3 and 30 eV," Phys. Stat. Sol. (b) **49**, 561-570 (1972).
[52] S. Tutihasi, G.G. Roberts, R.C. Keezer, and R.E. Drews, "Optical properties of tellurium in the fundamental absorption region," Phys. Rev. **177**(3), 1144-1150 (1969).
[53] J. Toudert, and R. Serna, "Ultraviolet-visible interband plasmonics with p-block elements," Opt. Mat. Expr. **6**(7), 2434-2447.
[54] S.Y. Huang, T.J. Kim, Y.W. Jung, N.S. Barange, H.G. Park, J.Y. Kim, Y.R. Kang, Y.D. Kim, S.H. Shin, J.D. Song, C.-T. Liang, Y.-C. Chang, "Dielectric function and critical points of AlP determined by spectroscopic ellipsometry," J. Alloys Comp. 587, 361-364 (2014).
[55] F. Meyer, E.E. de Kluizenaar, and D. den Engelsen, "Ellipsometric determination of the optical anisotropy of gallium selenide," J. Opt. Soc. Am **63**(3), 529-532 (1973).
[56] S. Adachi, *Handbook on physical properties of semiconductors* (Kluwer, 2004).
[57] J.-W. Park. M. Song, S. Yoon, H. Lim, D.S. Jeong, B.-K. Cheong, and H. Lee, "Stuctural and optical properties of phase-change amorphous and crystalline $Ge_{1-x}Te_x$ ($0 < x < 1$) thin films," Phys. Stat. Sol. A **210**(2), 265-275 (2013).
[58] S. Logothetidis, and H.M. Polatoglou, "Ellipsometric studies of the dielectric function of SnSe and a single model of the electronic structure and the bonds of the orthorhombic IV-VI compounds," Phys. Rev. B 36(14), 7491-7499 (1987).
[59] N. Suzuki, and S. Adachi, "Optical properties of SnTe," Jpn. J. Appl. Phys. **34**, 5977-5983 (1995).
[60] J.N. Zemel, J.D. Jensen, and R.B. Schoolar, "Electrical and optical properties of epitaxial films of PbS, PbSe, PbTe, and SnTe," Phys. Rev. 140(1), A330-A342 (1965).
[61] M. Eddrief, F. Vidal, and B. Gallas, "Optical properties of $Bi_2Se_3$: from bulk to ultrathin films," J. Phys. D: Appl. Phys. **49**, 505304 (2016).
[62] J. Humlicek, D. Hemzal, A. Dubroka, O. Caha, H. Steiner, G. Bauer, and G. Springholz, "Raman and interband optical spectra of epitaxial layers of the topological insulators $Bi_2Te_3$ and $Bi_2Se_3$ on $BaF_2$ substrates," Phys. Scr. **T162**, 014007 (2014).
[63] J. M. McMahon, G. C. Schatz, and S. K. Gray, "Plasmonics in the ultraviolet with the poor metals Al, Ga, In, Sn, Tl, Pb, and Bi," Phys. Chem. Chem. Phys. **15**(15), 5415–5423 (2013).

[64] K. Shportko, S. Kremers, M. Woda, D. Lencer, J. Robertson, and M. Wuttig, "Resonant bonding in crystalline phase-change materials," Nat. Mater. **7**(8), 653–658 (2008).
[65] A.M. Dubrovkin, G. Adamo, J. Yin, L. Wang, C. Soci, Q.J. Wang, and N.I. Zheludev, "Visible range plasmonic modes on topological insulator nanostructures," Adv. Optical Mater. **5**(3), (2017).
[66] J. Yin, H.N.S. Krishnamoorthy, G. Adamo, A.M. Dubrovkin, Y.D. Chong, N.I. Zheludev, and C. Soci, "Plasmonics of topological insulators at optical frequencies," **arXiv:1702.00302** (2017).
[67] P.I. Kuznetsov, G.G. Yakushcheva, B.S. Shchamkhalova, V.A. Luzanov, A.G. Temiryazev, and V.A. Jitov, "Metalorganic vapor phase epitaxy of ternary rhomboedral $(Bi_{1-x}Sb_x)_2Se_3$ solid solutions," J. Cryst. Growth **433**, 114-121 (2016).
[68] C.G. Das, R. Das, and D. Chakravorty, "Optical properties of bismuth granules in a glass matrix," Bull. Mater. Sci. **5**, 277 (1983).
[69] Y.W. Wang, B.H. Hong, and K.S. Kim, "Size control of semimetal bismuth nanoparticles and the UV-visible and IR absorption spectra", J. Phys. Chem. B **109**, 7067 (2005).
[70] S. Sivaramakrishnan, V.S: Muthukumar, S. Sivasankara Sai, K. Venkataramaniah, J. Reppert, A. Rao, M. Anija, R. Philip, N. Kuthirummal, "Nonlinear optical scattering and absorption in bismuth nanorod suspensions," Appl. Phys. Lett. **91**, 093104 (2007).
[71] X. Liu, Q. Luo, and J. Qiu, "Emerging low-dimensional materials for nonlinear optics and ultrafast photonics," Adv. Mater. **1605886** (2017).
[72] M. W. Knight, T. Coenen, Y. Yang, B. J. M. Brenny, M. Losurdo, A. S. Brown, H. O. Everitt, and A. Polman, "Gallium plasmonics: deep subwavelength spectroscopic imaging of single and interacting gallium nanoparticles," ACS Nano **9**(2), 2049–2060 (2015).
[73] F. Yang, N. Akozbek, T.-H. Kim, J.M. Sanz, F. Moreno, M. Losurdo, A.S. Brown, and H.O. Everitt, "Ultraviolet-visible plasmonic properties of gallium nanoparticles investigated by variable-angle spectroscopic and Mueller matrix ellipsometry," ACS Photonics **1**(7), 582-589 (2014).
[74] K. F. MacDonald and N. I. Zheludev, "Active plasmonics: current status," Laser Photonics Rev. **4**(4), 562–567 (2010).
[75] B. F. Soares, K. F. MacDonald, V. A. Fedotov, and N. I. Zheludev, "Light-induced switching between structural forms with different optical properties in a single gallium nanoparticulate," Nano Lett. **5**(10), 2104–2107 (2005).
[76] N. I. Zheludev, "Nonlinear optics on the nanoscale," Contemp. Phys. **43**(5), 365–377 (2002).
[77] S.R.C. Vivekchand, C.J. Engel, S.M. Lubin, M.G. Blaber, W. Zhou, J.Y. Suh, G.C. Schatz, and T.W. Odom, "Liquid plasmonics: manipulating surface plasmon polaritons via phase transitions," Nano Lett. **12**(8), 4324-4328 (2012).
[78] E. Haro-Poniatowski, R. Serna, M. Jiménez de Castro, A. Súarez-García, C.N. Afonso, and I. Vickridge, "Size-dependent thermos-optical properties of embedded Bi nanostructures," Nanotechnology **19**, 485708 (2008).
[79] T. Lewi, P.P. Iyier, N.A. Butakov, A.A. Mikhailovsky, and J.A. Schuller, "Widely tunable infrared antennas using free carrier refraction," Nano Lett. **15**(12), 8188-8193 (2015).
[80] S. Makarov, S. Kudryashov, I. Mukhin, A. Mozharov, V. Milichko, A. Krasnok, and P. Belov, "Tuning of magnetic optical response in a dielectric nanoparticle by ultrafast photoexcitation of dense electron-hole plasma," Nano Lett. **15**(9), 6187-6192 (2015).
[81] D.G. Baranov, S.V. Makarov, V.A. Milichko, S.I. Kudryashov, A.E. Krasnov, and P.A. Belov, "Nonlinear transient dynamics of photoexcited resonant silicon nanostructures," ACS Photonics **3**(9), 1546-1551 (2016).
[82] E.H. Sargent, "Infrared quantum dots," Adv. Mater. **17**(5) 515 – 521 (2005).
[83] T. Lewi, H.A. Evans, N.A. Butakov, and J.A. Schuller, "Thermo-optically reconfigurable PbTe metal-atoms," **arXiv:1703.00886** (2017).
[84] J. Hu, L. Li, H. Lin, Y. Zou, Q. Du, C. Smith, S. Novak, K. Richardson, and J.D. Musgraves, "Chalcogenide glass microphotonics: stepping into the spotlight," Am. Ceram. Soc. Bull. **94**(4) 24-29 (2015).
[85] Y. Zhou, D. Zhang, H. Lin, L. Li, L. Moreel, J. Zhou, Q. Du, O. Ogbuu, S. Danto, D. Musgraves, K. Richardson, K.D. Dobson, R. Birkmire, and J. Hu, "High-performance, high-index contrast chalcogenide glass photonics on silicon and unconventional non-planar substrates," Adv. Opt. Mater. **2**, 478-486 (2014).
[86] H. Lin, L. Li, Y. Zou, S. Danto, J.D. Musgraves, K. Richardson, S. Kozacik, M. Murakowski, D. Prather, P.T. Lin, V. Singh, A. Agarwal, L.C. Kimerling, and J. Hu, "Demonstration of high-Q mid-infrared chalcogenide glass-on-silicon resonators," Opt. Lett. **38**(9), 1470-1472 (2013).
[87] A. Titl, A.-K.U. Michel, M. Schäferling, X. Yin, B. Gholipour, L. Cui, M. Wuttig, T. Taubner, F. Neubrech, and H. Giessen, "A switchable mid-infrared plasmonic perfect absorber with multispectral thermal imaging capability," Adv. Mater. **27**, 4597-4603 (2015).
[88] A.-K.U. Michel, P. Zalden, D.N. Chigrin, M. Wuttig, A.M. Lindenberg, and T. Taubner, "Reversible optical switching of infrared antenna resonance with ultrathin phase-change layers using femtosecond laser pulses," ACS Photonics **1**, 833-839 (2014).
[89] A. Karvounis, B. Gholipour, F.F. MacDonald, and N.I. Zheludev, "All-dielectric phase-change reconfigurable metasurface," Appl. Phys. Lett. **109**, 051103 (2016).
[90] C.H. Chu, M.L. Tseng, J. Chen, P.C. Wu, Y.-H. Chen, H.-C. Wang, T.-Y. Chen, W.T. Hsieh, H.J. Wu, G. Sun, and D.P. Tsai, "Active dielectric metasurface based on phase-change medium," Lasers and Photon. News **10**(6), 986-994 (2016).

[91] M. Zhao, M. Bosman, M. Danesh, M. Zeng, P. Song, Y. Darma, A. Rusydi, H. Lin, C.-W. Qiu, and K.P. Loh, "Visible surface plasmon modes in single $Bi_2Te_3$ nanoplate," Nano Lett. **15**, 8331-8335 (2015).

[92] M. Zhao, J. Zhang, N. Gao, P. Song, M. Bosman, B. Peng, B. Sun, C.-W. Qiu, Q.-H. Xu, Q. Bao and K.P. Loh, "Actively tunable visible surface plasmons in $Bi_2Te_3$ and their energy-harvesting applications," Adv. Mater. **28**, 3138-3144 (2016).

[93] J. Guozhi, W. Peng, Z. Yanbang, and C. Kai, "Localized surface plasmon enhanced photothermal conversion in $Bi_2Se_3$ topological insulator nanoflowers," Sci. Rep. **6**, 25884 (2016).

[94] A. Vargas, S. Basak, F. Liu, B. Wang, E. Panaitecu, H. Lin, R. Markiewicz, A. Bansil, and S. Kar, "The changing colors of a quantum-confined topological insulator," ACS Nano **8**(2), 1222-1230 (2014).

[95] A. Vargas, F. Liu, and S. Kar, "Giant enhancement of light emission from nanoscale $Bi_2Se_3$", Appl. Phys. Lett. **106**, 243107 (2015).

[96] J.J. Cha, K.J. Koski, K.C.Y. Huang, K.X. Wang, W. Luo, D. Kong, Z. Yu, S. Fan, M.L. Brongersma, and Y. Cui, "Two-dimensional chalcogenide nanoplates as tunable metamaterials via chemical intercalation," Nano Lett. **13**(12), 5913-5918 (2013).

[97] W.S. Whitney, V.W. Brar, Y. Ou, Y. Shao, A.R. Davoyan, D.N. Basov, K. He, Q.-K. Xue, and H.A. Atwater, "Gate-variable mid-infrared optical transitions in a $(Bi_{1-x}Sb_x)_2Te_3$ topological insulator," Nano Lett. **17**(1), 255-260 (2017).

[98] M. Esslinger, R. Vogelsgesang, N. Talebi, W. Khunsin, P. Gehring, S. de Zuani, B. Gompf, and K. Kern, "Tetradymites as natural hyperbolic materials for the near-infrared to visible," ACS Photonics **1**, 1285-1289 (2014).

[99] N. Talebi, C. Ozsoy-Keskinbora, H.M. Benia, K. Kern, C.T. Koch, and P.A. van Aken, "Wedge Dyakonov waves and Dyakonov plasmons in topological insulator of $Bi_2Se_3$ probed by electron beams," ACS Nano **10** (7), 6988–6994 (2016).

[100] J. Sun, N.M. Litchinitser, and J. Zhou, "Indefinite by nature: from ultraviolet to terahertz," ACS Photonics **1**, 293-303 (2014).

[101] K. Korzeb, M. Gajc, and D. A. Pawlak, "Compendium of natural hyperbolic materials," Opt. Express **23**(20), 25406–25424 (2015).

[102] H. Cui, I.B. Bhat, and R. Venkatasubramanian, "Optical constants of $Bi_2Te_3$ and $Sb_2Te_3$ measured using spectroscopic ellipsometry," J. Electron. Mater. **28**(10), 1111-1114 (1999).

[103] A. Green, S. Dey, Y.Q. An, B. O'Brien, S. O'Mullane, B. Thiel, and A.C. Diebold, "Surface oxidation of the topological insulator $Bi_2Se_3$," J. Vac. Sci. Technol. A **34**(6), 061403 (2016).

[104] D.L. Greenaway and G. Harbeke, "Band structure of bismuth telluride, bismuth selenide and their respective alloys," J. Phys. Chem. Sol. **26**, 1585-1604 (1965).

[105] H. Peng, W. Dang, J. Cao, Y. Chen., D. Wu, W. Zheng, H. Li, Z.-X. Shen, and Z. Liu, "Topological insulator nanostructures for near-infrared transparent flexible electrodes," Nat. Chem. **4**, 281-286 (2012).

[106] J. Yan, Z. Lin, C. Ma, Z. Zheng, P. Liu, and G. Yang, "Plasmon resonances in semiconductor materials for detecting photocatalysis at single-particle level," Nanoscale **8**, 15001-15007 (2016).

[107] L. Braic, N. Vasilantonakis, A.P. Mihai, I.J. Villar Garcia, Sarah Fearn, B. You, B. Doiron, R.F. Oulton, L. Cohen, S.A. Maier, N. McN. Alford, A.V. Zayats, and P.K. Petrov, "Titanium oxynitride thin films with tunable double epsilon-near-zero-behaviour," **arXiv:1703.09467** (2017).

[108] S. De Zuani, M. Rommel, B. Gompf, A. Berrier, J. Weis, and M. Dressel, "Suppressed percolation in nearly closed gold films," ACS Photonics 3(6), 1109-1115 (2016).

**Figure 1**

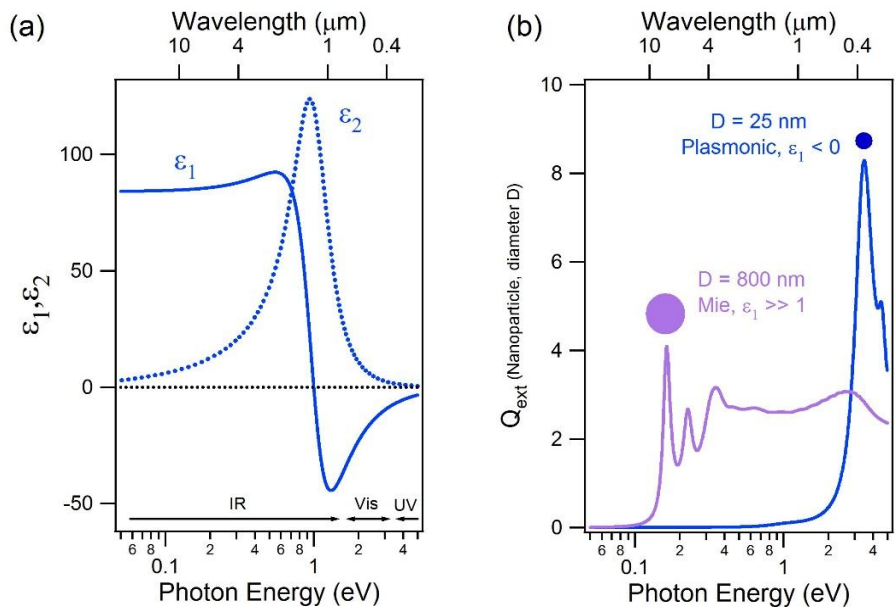

*Strong interband transitions drive both plasmonic and Mie resonances. (a) Real part ($\varepsilon_1$, continuous line) and imaginary part ($\varepsilon_2$, dotted line) of a Kramers-Kronig consistent dielectric function consisting of a single Lorentz oscillator (simulated using the WVASE32 software from J.A. Woollam Co.). (b) Optical extinction efficiency $Q_{ext}$ of spherical nanoparticles with the dielectric function shown in (a), for different diameters D, in a transparent medium with refractive index n = 1.5 (Simulated using the MiePlot software [20]).*

**Figure 2**

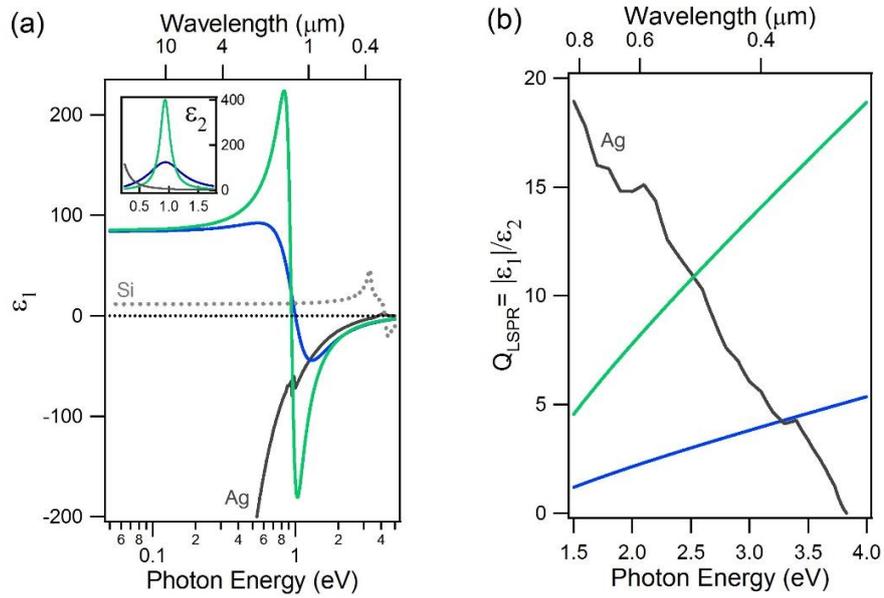

*Role of the interband transition amplitude and width on the plasmonic and Mie performance. (a) Real part $\varepsilon_1$ of Kramers-Kronig consistent dielectric functions consisting of a broad (blue line) and sharp (green line) Lorentz oscillators. The corresponding imaginary parts $\varepsilon_2$ are shown in the inset. The real part of the dielectric function of Ag [21] and Si [22] are shown for comparison. (b) Localized surface plasmon resonance quality factor $Q_{LSPR}$ [23, 24] for the dielectric functions of the broad and sharp oscillators (same colors as in (a)) and for Ag.*

**Figure 3**

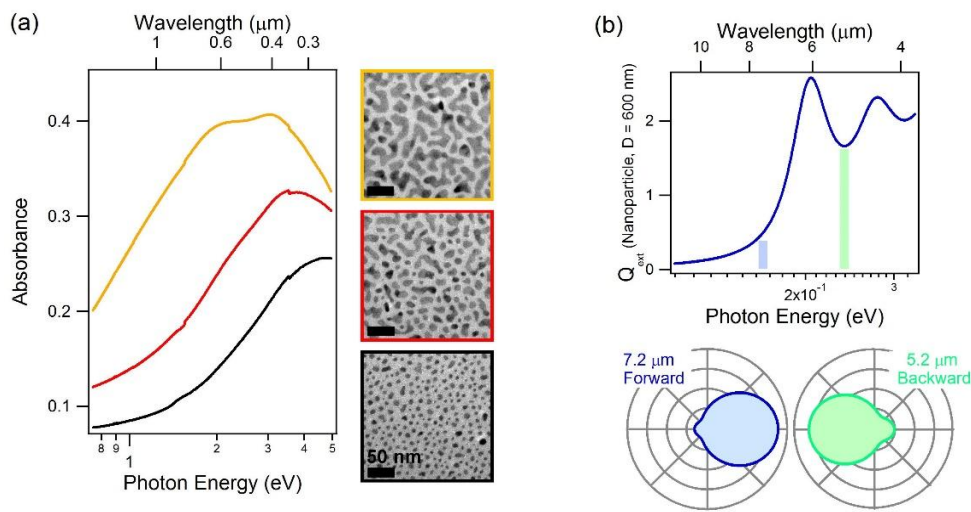

(*a*) *Experimental ultraviolet – visible – near infrared plasmonic resonances of Bi nanoparticles of different sizes and shapes embedded in transparent matrix (adapted from [14]). The images shown in the insets correspond to the optical absorbance spectra with the same colors. (b) Simulated mid infrared Mie resonances of a spherical Bi nanoparticle embedded in a transparent matrix. Its optical extinction efficiency spectrum ($Q_{ext}$) was calculated for a nanoparticle diameter D = 600 nm. The dielectric function $\varepsilon$ of Bi was taken from [25]), and that of the matrix was $\varepsilon_m$ = 2.72. The simulated scattering patterns of this nanoparticle are also shown, at photon wavelengths in vacuo of 5.2 μm and 7.2 μm, for unpolarized incident plane waves traveling from the left to the right.*

**Figure 4**

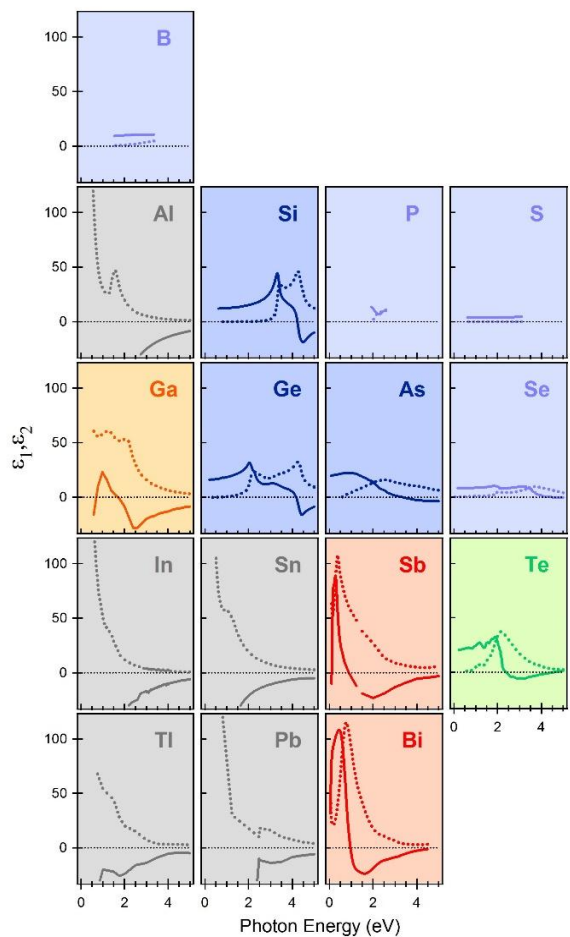

*Bulk dielectric functions ($\varepsilon_1$ solid line, $\varepsilon_2$ dashed line) of the elemental p-block materials, taken from the literature: B [37], Al [21], Ga [38], In [39,40], Tl [41], Si [42], Ge [43], Sn [21], Pb [44], P (black phosphorus, flakes) [45], As (amorphous) [46], Sb [47,48], Bi film similar to those studied in [25], S [49], Se [50,51], Te [52]. The most relevant elements for the study of interband plasmonic and Mie resonances are those represented in red, orange, red, green and blue (see text).*

**Figure 5**

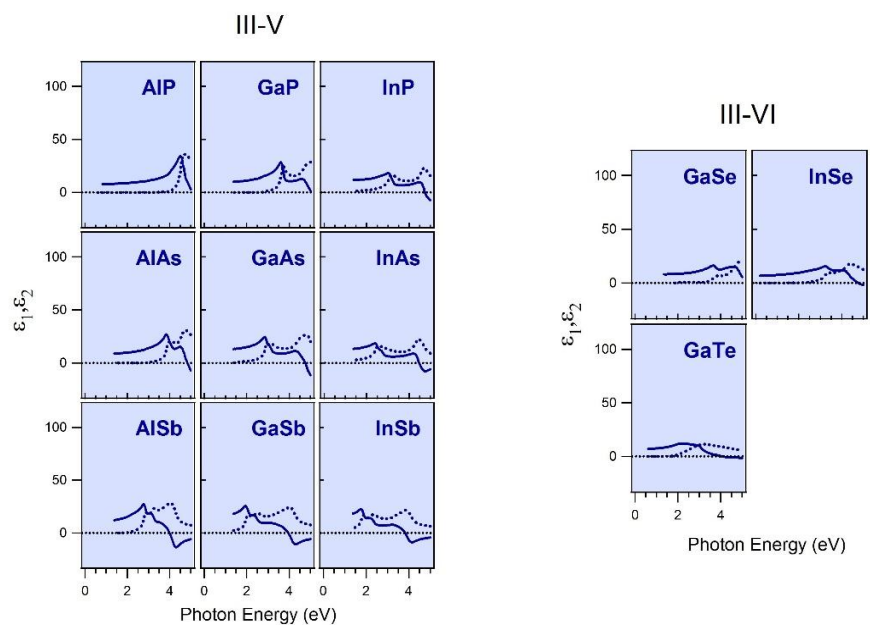

*Bulk dielectric functions ($\varepsilon_1$ solid line, $\varepsilon_2$ dashed line) of some binary compounds of p-block elements, taken from the literature: III-V compounds (AlP [54], AlAs, AlSb, GaP, GaAs, GaSb, InP, InAs, and InSb [43]) and III-VI chalcogenide compounds (GaSe [55], GaTe [56], InSe [56])*

**Figure 6**

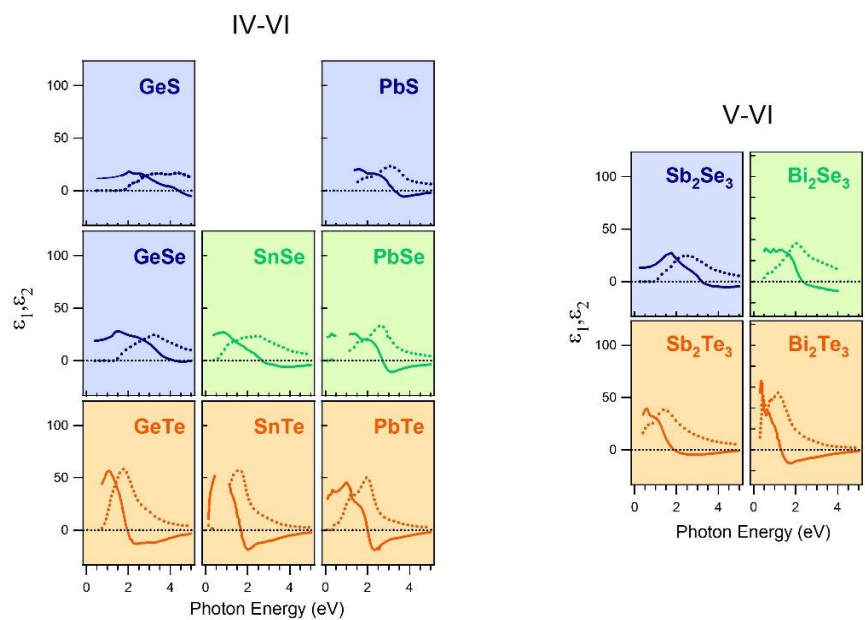

*Bulk dielectric functions ($\varepsilon_1$ solid line, $\varepsilon_2$ dashed line) of some binary compounds of p-block elements, taken from the literature: IV-VI chalcogenide compounds (GeS [56], GeSe [56], GeTe [57], SnSe [58], SnTe [56, 59], PbS [43, 60], PbSe [56], PbTe [56]); V-VI chalcogenide compounds ($Sb_2Se_3$ [56], $Sb_2Te_3$ [56], $Bi_2Se_3$ [61], $Bi_2Te_3$ [62]).*

**Figure 7**

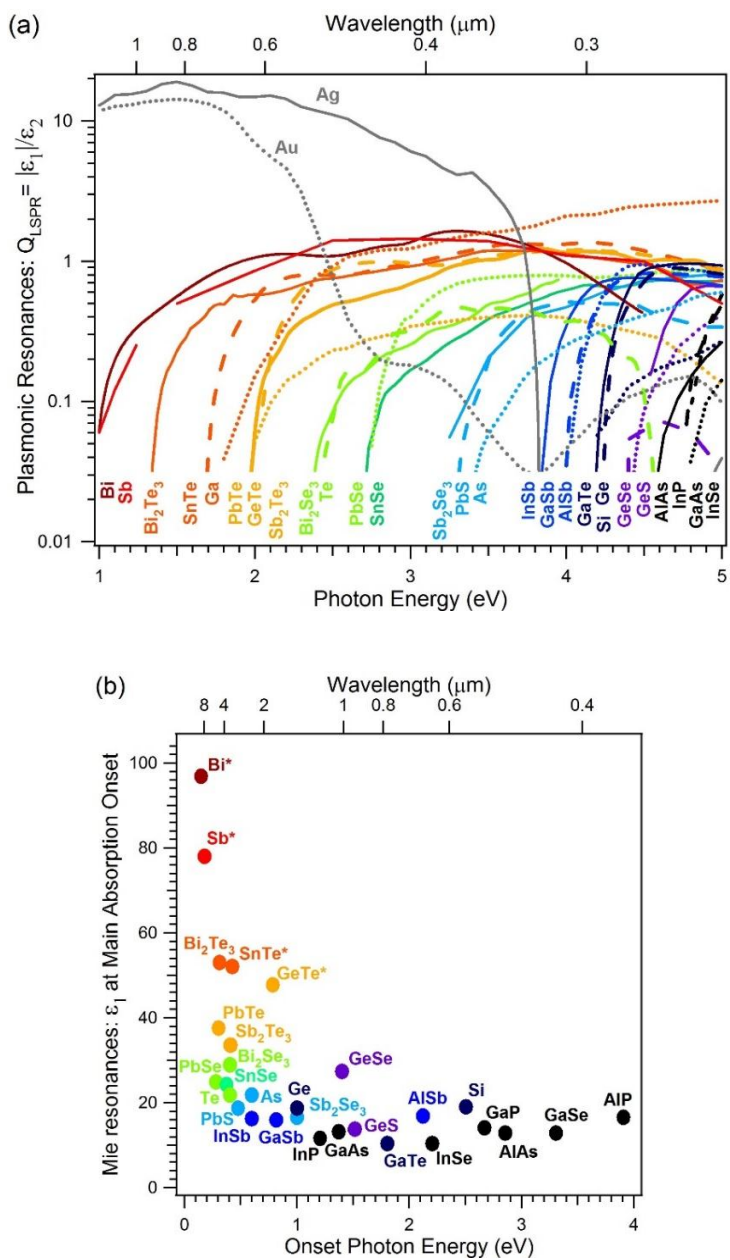

*Performance of selected single-elements and compounds of the p-block for interband plasmonic and Mie resonances: (a) Plasmonic quality factor $Q_{LSPR}$ vs photon energy in the ultraviolet to near infrared region. The quality factors of Au and Ag are also shown. (b) $\varepsilon_1$ at the main onset of optical absorption vs the onset photon energy for semiconductors and topological insulators (note: the onset is related with the main interband transition band, it is not the bandgap). For semi-metals, $\varepsilon_1$ was taken at the photon energy for which $\varepsilon_1$ is maximum and $\varepsilon_2$ is minimum. All the calculations were done using the dielectric functions shown in Figs. 4,5,6.*